# A Silicon cluster based single electron transistor with potential room temperature switching

Zhanbin Bai(白占斌)[1], Xiangkai Liu (刘翔凯)[2], Zhen Lian (连震)[3], KangKang Zhang(张康康)[1], Guanghou Wang (王广厚)[1], Su-Fei Shi(史夙飞)[3], Xiaodong Pi(皮孝东)[2], Fengqi Song(宋凤麒)[1,**]

[1] National Laboratory of Solid State Microstructures, School of Physics and Collaborative Innovation Center of Advanced Microstructures, Nanjing University, Nanjing 210093, China

[2] State Key Laboratory of Silicon Materials and School of Materials Science and Engineering, Zhejiang University, Hangzhou 310027 , China

[3] Department of Chemical and Biological Engineering, Rensselaer Polytechnic Institute, NY 12180, USA

**Supported by** the National Key Research and Development Program of China under Grant No 2017YFA0303200, the National Natural Science Foundation of China under Grant Nos U1732273, U1732159, 91421109, 91622115, 11522432, 11574217 and 61774133, the Natural Science Foundation of Jiangsu Province under Grant No BK20160659.

** Corresponding author. Email: songfengqi@nju.edu.cn. Fax: +86-25-83595535

%Email: 605763236@qq.com; xiangkailiu@163.com; lianz@rpi.edu;

  kkzhang@smail.nju.edu.cn; shis2@rpi.edu; xdpi@zju.edu.cn.

%Tel. 13813905754     15251897909

**Abstract**

We demonstrate the fabrication of a single electron transistor device based on a single ultra-small silicon quantum dot connected to a gold break junction with a nanometer scale separation. The gold break junction is created through a controllable electromigration process and the individual silicon quantum dot in the junction is deter-mined to be a Si170 cluster. Differential conductance as a function of the bias and gate voltage clearly shows the Coulomb diamond which confirms that the transport is dominated by a single silicon quantum dot. It is found that the charging energy can be as large as 300 meV, which is a result of the large capacitance of a small silicon quantum dot (∼1.8 nm). This large Coulomb interaction can potentially enable a single electron transistor to work at room temperature. The level spacing of the excited state can be as large as 10 meV, which enables us to manipulate individual spin via an external magnetic field. The resulting Zeeman splitting is measured and the g factor of 2.3 is obtained, suggesting relatively weak electron-electron interaction in the silicon quantum dot which is beneficial for spin coherence time.

The single electron transistor (SET) is a promising component in future nanodevice-constructed computers, where the electrons inside the device repels the electron from the source electrode with the result of Coulomb blockade and single electron switching due to quantum size constriction in various devices such as lithographically defined semiconducting quantum dots (QDs) [1,2], single molecules[3-5], or metal nanoparticles[6-8]. By tuning the gate, the chemical potential of the dot will be modulated continuously, and these discrete energy levels will emerge one by one, showing coulomb oscillation. The silicon-based SETs have attracted intense attention due to their compatibility to modern semiconductor processing and lower spin relaxation rate due to their weak spin-orbit coupling and zero nuclear spin. Single-shot readout of an electron spin in silicon is first performed experimentally in a device consisting of implanted phosphorus donors coupled to a metal-oxide-semiconductor single-electron transistor, with a spin lifetime of ~6 s at a magnetic field of 1.5 tesla[9].

The charging energy $E_c$ is one of its central parameters in seeking the room-temperature working SET, which is normally determined by the dot size in given materials. The early value is around 1.5 meV for a large single-electron quantum dot lithographically defined in two-dimensional electron gas[10]. It later evolves to 6 meV due to smaller dimension and advanced fabrication techniques[11]. The bottom-up synthesis of solution-processable Si QDs has made great progress in recent years. The mean size of solution-processable Si QDs can now be routinely tuned from ~10 to ~2 nm[12-17]. Thus, Si QDs that are small enough to be called Si clusters may exist in an ensemble of solution-processable Si QDs with a small mean size. On the other hand, the break-junction technique[18,19] has been well developed with the result of a reliable solution-compatible single molecular device technique[20]. These shed light on the advanced SET devices based on the solution-processable Si clusters.

In this Letter, we report the successful fabrication of the nanodevices with individual Si QDs. The trans-port measurements give typical SET behavior of the Coulomb blockade and oscillations. Its charging energy reaches over 300 meV, indicating a very large Ec and possible room-temperature working. Its Zeeman effect is also studied. The captured QD is determined to be a $Si_{170\pm15}$ cluster. This is the smallest Si SET device based on atomic clusters, to our best knowledge.

The atomic cluster is a very small QD captured from the toluene solution of Si QDs prepared in a nonthermal plasma and subsequently hydrosilylated with 1-octene[13]. In order to measure this ultra-small Si QD, we utilize electromigration[18] to form a three-terminal transistor. Firstly, e-beam lithography is used to pattern a 70-nm-wide electrode, and Au is deposited by EBE[19]. All of these narrow Au electrodes are located on a 30-nm-thick local silicon dioxide layer as back gates. After the toluene solution of Si QDs is dropped on the wafer and dries out, the device is cooled down to liquid helium temperature in a superconducting magnet system from cryomagnetics. Using a Keithley 6430 source meter, we perform electromigration to make the break junctions for a gold electrode. The atomic force microscopic (AFM) image is taken by the Cypher from Oxford instrument.

Figures 1(a) and 1(b) show the TEM image of Si QDs and their size distribution, respectively. The Si QDs are with the mean size of 2.7 nm from the histogram statistics. As shown in Fig. 1(c), the efficient photoluminescence of Si QDs peaks at the wavelength of about 695 nm, indicating the high quality of Si QDs used in the current work. Please note that Si QDs are nearly spherical[21], allowing the estimation of the atomic quantity by the diameter. There are some ultrafine QDs with the diameters of even smaller than 2 nm, which is reasonable due to some inhomogeneity of the QD synthesis. What we captured in this work is such an atomic scale Si cluster.

We utilize the electromigration technique to form the break junction. We apply a small voltage to the e-beam lithographically defined 70-nm-wide Au wire electrode (Fig. 1(d)), which has been immersed in the toluene solution with Si QDs beforehand. The applied voltage will drive the electromigration of the gold atoms near some weak point on the Au nanowire and finally break the wire with a nanosized gap. Simultaneously the Si QD will fall in the gap with the result of a three-terminal junction. The feedback-controlled breaking process[22,23] is shown in Fig. 1(e), where in a typical turn we increase the bias voltage until the current drops, then decrease the voltage back to 0.1 V immediately and increase again. Finally after the resistance is more than about 1 MΩ[24], a 1~2-nm-wide gap is believed to be formed[3,19]. Figure 1(f) shows the morphology feature taken by an atomic force microscope (AFM) after the measurement. It is not very clear due to the residual solution while we can still see the captured Si QD with enhanced contrast. Reason-ably, the assumed narrow gap is found to be broadened to several nanometers because of high mobility of gold atoms under room-temperature exposure[25].

Due to the large resistance, we carry out the dc I-V[26] measurements using a Keithley 2450 source meter to provide gate voltage and a Keithley 6430 source meter to provide bias voltage and measure bias current[27]. We then calculate the numerical differential conductance. Figure 2(a) presents the measured I-V curves while several different gate voltages are applied at T=4 K, which show typical coulomb blockade effect with a large charging energy of up to 300 meV. Figure 2(b) shows the source-drain current $I_{sd}$ for fixing a constant bias voltage of 2 mV and for sweeping the gate voltage from −10 V to 7 V. It is unexpected that only one coulomb peak appears in the whole gate regime. The degeneracy point is found near the gate voltage of −4.6 V.

We perform the detailed spectroscopic measurements of the single Si QD at T= 1.6 K by recording the current when sweeping bias voltage from 50 mV to −50 mV, and repeat it by stepping the values of gate voltage $V_g$ around the degeneracy point. Then differential conductance $dI / dV_{sd}$ is plot versus g and sd as shown in Fig. 2(c). It must be noted that the degeneracy point moves a little along g at different temperatures. The QD is found to be weakly coupled[28,29] to the gold electrodes since the differential conductance peak is very small but still distinguishable. There is a brighter line parallel to the coulomb edge, indicating a nearby excited state at about 10 mV under degeneracy point. However, the line becomes invisible when applying positive bias voltages. The spectroscopic asymmetry originates from the asymmetry of the left and right tunneling barriers[30] and such asymmetric electronic coupling quite commonly appears in break junctions[31,32]. The degeneracy point separates two regimes with different electron numbers. The switching ratio is well over 10. This confirms the formation of the Si QD SET with a large $E_c$ in our experiment.

The size of the QDs can be more precisely determined by analyzing Fig 2(c). According to the position of the excited state or the slopes of two coulomb edges, we can derive that the coupling parameter α between Si QD and $SiO_2$ back gate layer is about 0.02, indicating that a very weak coupling between them. In our device, an unexpected phenomenon is that energy spacing is out of modulation range as shown in Fig. 2(b). We cannot clearly see any more feature when the bias voltage reachs the upper energy level even in another rough differential conductance counter plot with the bias voltage of up to 300mV and the gate voltage of up to 10V. This means we only find one degeneracy point between N and N-1 electrons. This becomes rather reasonable when we remember that the smaller the silicon QD is, the larger $E_c$ is. This has been evident in the recent study of Silicon QDs. A 4.3nm-diameter silicon QD gives the $E_c$ of 56meV[16] and Another 10nm-diameter Si QD gives the $E_c$ of 11meV[17]. Fitting the previous data by using the charging energy of about 300meV, we can determine that an ultra-small silicon dot of 1.8±0.1nm diameter is located in the electromigration-formed gap, which results in the SET devices. Estimating from the crystalline structure of bulk silicon, we are convinced that the measured device contains a silicon atomic cluster with 170±15 atoms. The very small size of the atomic cluster allows the possible room-temperature working of the Si SET devices.

The Zeeman effect can be induced and measured in the SET devices when we apply a perpendicular magnetic field from 8T to -4T as shown in Figure 3(a)~(d). The degeneracy point shifts a little as the magnetic field changes. This is known as "global shift" and common in semiconducting QD differential conductance measurement[33]. It is to be noted in the data that the linewidth of the two coulomb edges on the right side of degeneracy point decrease monotonously as the magnetic field decrease from 8T to -4T[34]. Here we take the conductance data of one edge on the upper-right side of degeneracy point to statistically obtain the change of the linewidth. We shift bias voltage of each $dI/dV$ curve according to the value of its corresponding gate voltage Vg so that conductance peak on all gate voltage can almost be aligned at the same level of bias voltage. The normalized $dI/dV$ curves are shown in fig 3(e), and the FWHM of all dI/dV peaks are extracted and shown in Fig. 3(f) with an approximate slope of 0.3±0.025mV/T. This can calculates a Lande g-factor of 2.3 with an uncertainty of 0.35, although we note that there might be some other physics. It is reasonable that the g-factor is close to 2 since no complex interaction is desired in Si QDs.

As a perspective, the study on the atomic Si clusters may introduce a new spectroscopy method in the cluster sciences. As long known, the cluster science communities rely on the photoemission spectroscopy to study the electronic structures of the atomic cluster[35-38], which finally gives the atomic structures of the clusters by fitting the data using the first principal calculations. It suffers from low resolution of around 0.1eV due to uncontrolled cluster movements during the measurements. The SET study of the atomic cluster may resolve the electronic states of the atomic clusters in a meV-order resolution and some vibrational modes, which therefore allows in-depth insight of the atomic cluster studies. The Zeeman effect study on different energy levels also allows detailed understanding of the electronic interactions.

In summary, using the electromigration break junction, we have demonstrated a 170±15-atom silicon cluster based SET under weak-coupling condition. The large charging energy may forebode a robust stability of working performance under room temperature. This paves the application of the Si SET in future nanodevice computers.

We acknowledge the helpful assistance of the Nanofabrication and Characterization Center at the Physics College of Nanjing University.

**References**

[1] Kouwenhoven L P, Oosterkamp T H, Danoesastro M W S, Eto M, Austing D G, Honda T, and Tarucha S 1997 Science **278** 1788.

[2] Hanson R, Kouwenhoven L P, Petta J R, Tarucha S, and Vandersypen L M K 2007 Rev. Mod. Phys. **79** 1217.

[3] Park J, Pasupathy A N, Goldsmith J I, Chang C, Yaish Y, Petta J R, Rinkoski M, Sethna J P, Abruña H D, McEuen P L, and Ralph D C 2002 Nature **417** 722.

[4] Liang W, Shores M P, Bockrath M, Long J R, and Park H 2002 Nature **417** 725.

[5] Park H, Park J, Lim A K L, Anderson E H, Alivisatos A P, and McEuen P L 2000 Nature **407** 57.

[6] Kuemmeth F, Bolotin K I, Shi S-F, and Ralph D C 2008 Nano Lett. **8** 4506.

[7] Bolotin K I, Kuemmeth F, Pasupathy A N, and Ralph D C 2004 Appl. Phys. Lett. **84** 3154.

[8] Petta J R and Ralph D C 2001 Phys. Rev. Lett. **87** 266801.

[9] Morello A, Pla J J, Zwanenburg F A, Chan K W, Tan K Y, Huebl H, Mottonen M, Nugroho C D, Yang C, van Donkelaar J A, Alves A D, Jamieson D N, Escott C C, Hollenberg L C, Clark R G, and Dzurak A S 2010 Nature **467** 687.

[10] Simmons C B, Thalakulam M, Shaji N, Klein L J, Qin H, Blick R H, Savage D E, Lagally M G, Coppersmith S N, and Eriksson M A 2007 Appl. Phys. Lett. **91** 213103.

[11] Lim W H, Zwanenburg F A, Huebl H, Möttönen M, Chan K W, Morello A, and Dzurak A S 2009 Appl. Phys. Lett. **95** 242102.

[12] Liu X, Zhang Y, Yu T, Qiao X, Gresback R, Pi X, and Yang D 2016 Part. Part. Syst. Charact. **33** 44.

[13] Liu X, Zhao S, Gu W, Zhang Y, Qiao X, Ni Z, Pi X, and Yang D 2018 ACS Appl. Mater. Interfaces **10** 5959.

[14] Gu W, Liu X, Pi X, Dai X, Zhao S, Yao L, Li D, Jin Y, Xu M, Yang D, and Qin G 2017 IEEE Photon. J. **9** 1.

[15] Yu T, Wang F, Xu Y, Ma L, Pi X, and Yang D 2016 Adv Mater **28** 4912.

[16] Zaknoon B, Bahir G, Saguy C, Edrei R, Hoffman A, Rao R A, Muralidhar R, and Chang K-M 2008 Nano Lett. **8** 1689.

[17] Sawada T, Kodera T, and Oda S 2016 Appl. Phys. Lett. **109** 213102.

[18] Park H, Lim A K L, Alivisatos A P, Park J, and McEuen P L 1999 Appl. Phys. Lett. **75** 301.

[19] Shi S F, Xu X, Ralph D C, and McEuen P L 2011 Nano Lett. **11** 1814.

[20] Heersche H B, de Groot Z, Folk J A, van der Zant H S, Romeike C, Wegewijs M R, Zobbi L, Barreca D, Tondello E, and Cornia A 2006 Phys. Rev. Lett. **96** 206801.

[21] Young N P, Li Z Y, Chen Y, Palomba S, Di Vece M, and Palmer R E 2008 Phys. Rev. Lett. **101** 246103.

[22] Strachan D R, Smith D E, Johnston D E, Park T H, Therien M J, Bonnell D A, and Johnson A T 2005 Appl. Phys. Lett. **86** 043109.

[23] Houck A A, Labaziewicz J, Chan E K, Folk J A, and Chuang I L 2005 Nano Lett. **5** 1685.

[24] Xu B and Tao N J 2003 Science **301** 1221.

[25] O'Neill K, Osorio E A, and van der Zant H S J 2007 Appl. Phys. Lett. **90** 133109.

[26] Jo M-H, Grose J E, Baheti K, Deshmukh M M, Sokol J J, Rumberger E M, Hendrickson D N, Long J R, Park H, and Ralph D C 2006 Nano Lett. **6** 2014.

[27] Garrigues A R, Wang L, del Barco E, and Nijhuis C A 2016 Nature Commun **7**

[28] Frisenda R and van der Zant H S J 2016 Phys. Rev. Lett. **117**

[29] Xiang D, Wang X, Jia C, Lee T, and Guo X 2016 Chem Rev **116** 4318.

[30] Kouwenhoven L P, Austing D G, and Tarucha S 2001 Rep. Prog. Phys. **64** 701.

[31] Parks J J, Champagne A R, Hutchison G R, Flores-Torres S, Abruna H D, and Ralph D C 2007 Phys. Rev. Lett. **99** 026601.

[32] Island J O, Gaudenzi R, de Bruijckere J, Burzuri E, Franco C, Mas-Torrent M, Rovira C, Veciana J, Klapwijk T M, Aguado R, and van der Zant H S 2017 Phys. Rev. Lett. **118** 117001.

[33] Weis J, Haug R J, Klitzing K v, and Ploog K 1993 Phys. Rev. Lett. **71** 4019.

[34] Grose J E, Tam E S, Timm C, Scheloske M, Ulgut B, Parks J J, Abruna H D, Harneit W, and Ralph D C 2008 Nature Mater **7** 884.

[35] Li J, Li X, Zhai H-J, and Wang L-S 2003 Science **299** 864.

[36] Bartels C, Hock C, Huwer J, Kuhnen R, Schwöbel J, and von Issendorff B 2009 Science **323** 1323.

[37] Zhai H J, Zhao Y F, Li W L, Chen Q, Bai H, Hu H S, Piazza Z A, Tian W J, Lu H G, Wu Y B, Mu Y W, Wei G F, Liu Z P, Li J, Li S D, and Wang L S 2014 Nature Chem **6** 727.

[38] Piazza Z A, Hu H S, Li W L, Zhao Y F, Li J, and Wang L S 2014 Nature Commun

**5** 3113.

**Figure captions**

**Figure 1 Construction of Si-QD SET.** (a) the high-resolution TEM image of the Si QDs. (b) The size distribution of Si QDs. (c) Optical absorption and PL spectra of Si QDs. (d) SEM image of nanowire electrode for electromigration. The inset shows the wafer on a chip carrier. (e) The feedback-controlled electromigration process. (f) The AFM image of Au nanowire after the break-junction measurement. It is not very clear because of the Si QD deposition.

**Figure 2** Si QD SET spectroscopy.(a) I-V curves of Si-QD SET on different gate voltage. (b) Isd-Vg on $V_{bias}$=2mV (DC measurement). (c) A more precise measurement around degeneracy point of the N-1 and N electron at T=1.6K. The conductance is numerically calculated from DC I-V curves because of its small magnitude ranging from 0 to 2nS. A clear N-1 excitation state is seen about 10mV under N ground state. The coupling strength between the QD energy level and the lead is asymmetric at positive and negative bias voltage.

**Figure 3** Transport under perpendicular magnetic field. (a)-(d) $dI/dV$ dependence on $V_{bias}$ and $V_g$ at B taken from -4T to 8T. The linewidth of these coulomb edges change with magnetic field out of plane. (e) The upper-right coulomb edge of the $dI/dV$ map broadens as magnetic field increase. (f) The FWHM of di/dv peak changes as magnetic field. We can derive that the *g* factor is about 2.3.

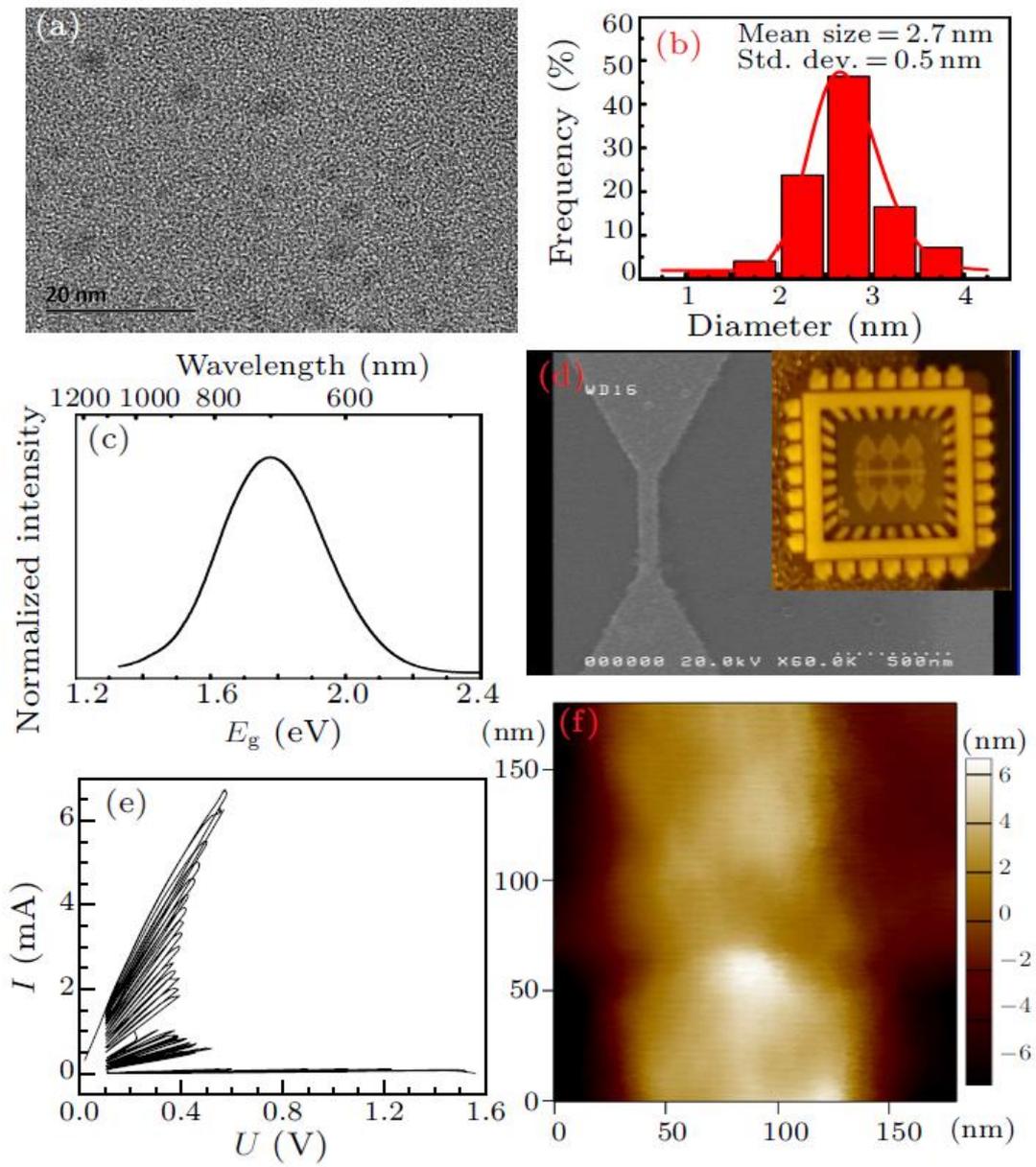

**Figure 1 Bai et al**

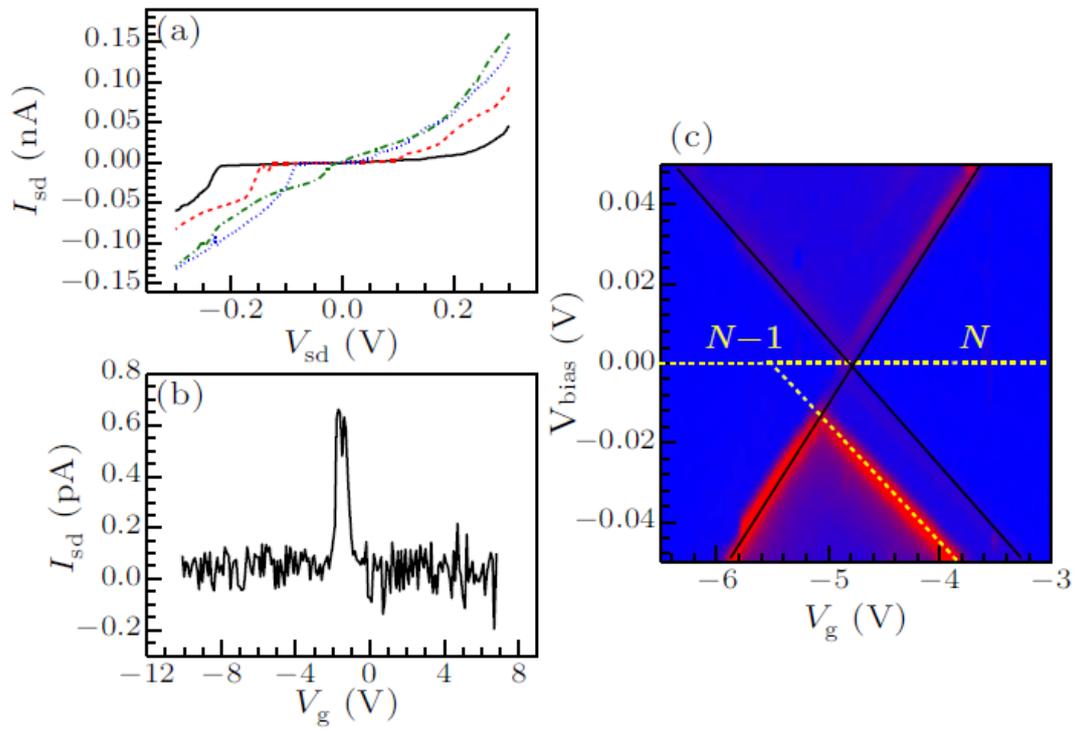

**Figure 2 Bai et al**

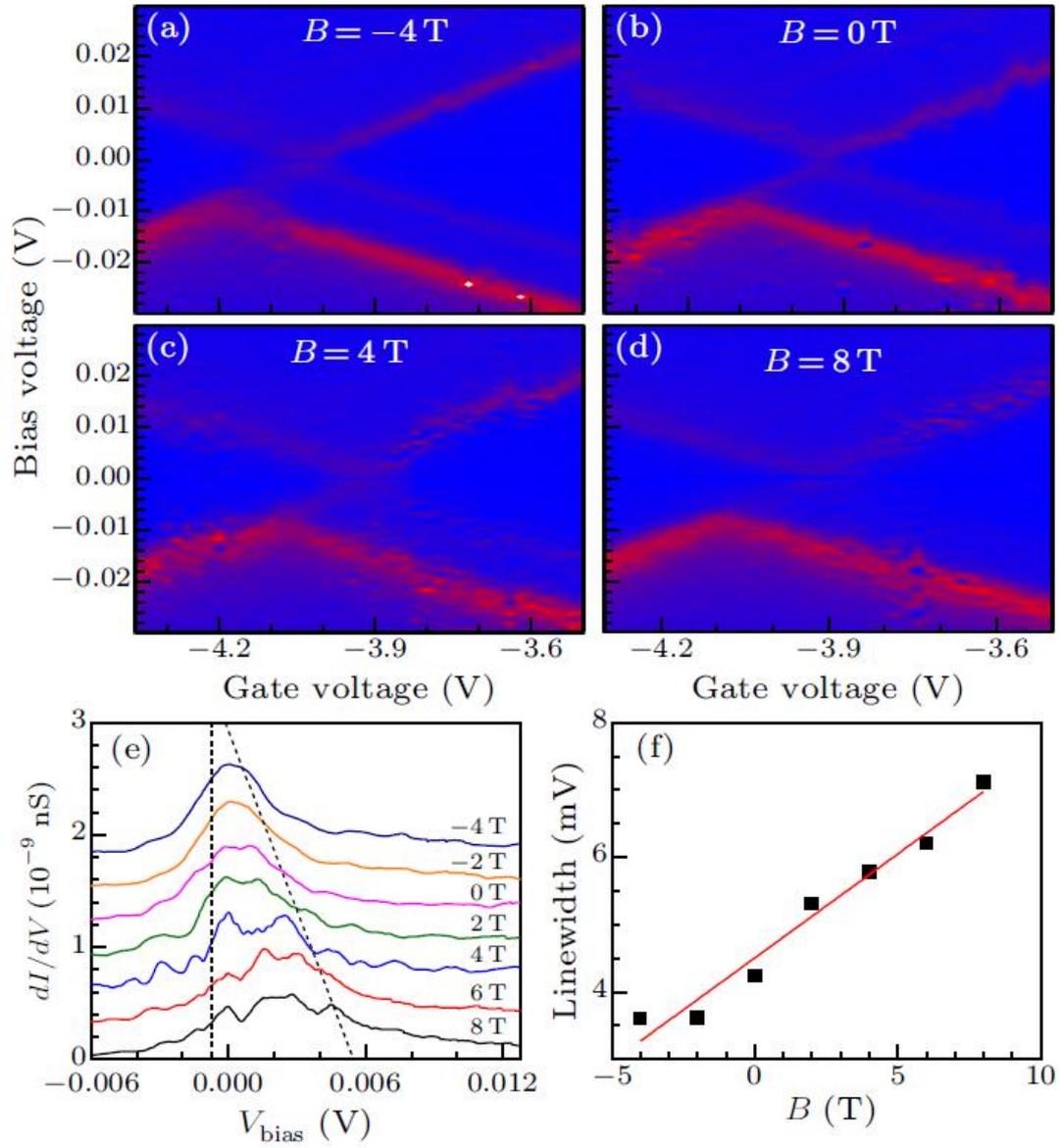

**Figure 3 Bai et al**
17